\documentclass[aps,prd,twocolumn,showpacs,floatfix,superscriptaddress,preprintnumbers,nofootinbib]{revtex4-1}
\usepackage{graphicx,amssymb}

\usepackage[english]{babel}
\usepackage{graphicx}
\usepackage{latexsym}
\usepackage{amssymb}
\usepackage{amsmath}

\begin{document}

\title[Non-orientable wormholes as portals to the mirror world]{Non-orientable wormholes
as portals to the mirror world}
\author{V.I. Dokuchaev}\thanks{e-mail: dokuchaev@inr.ac.ru}
\affiliation{Institute for Nuclear Research of the Russian Academy of
Sciences \\ 60th October Anniversary Prospect 7a, 117312 Moscow, Russia}
\author{Yu.N. Eroshenko}\thanks{e-mail: eroshenko@inr.ac.ru}
\affiliation{Institute for Nuclear Research of the Russian Academy of
Sciences \\ 60th October Anniversary Prospect 7a, 117312 Moscow, Russia}

\date{\today}

\begin{abstract}
We examine the compatibility of the mirror matter concept with the non-orientable wormholes. If
any particle (or classical object) is traversing through the non-orientable wormhole, it turns
into a corresponding mirror particle and vice versa. The critical sonic point in the hydrodynamical flow of the perfect fluid is located exactly 
at the throat of the wormhole. This means that wormholes are traversable for the fluid.
The astrophysical signatures of the presence of non-orientable wormholes in the Universe are discussed. Some non-trivial aspects of
the non-orientable wormholes electrodynamics are considered.
\end{abstract}

\maketitle


\section{Introduction}

The concept of mirror matter was proposed in the famous article of Lee and Yang
\cite{LeeYan56}, where they formulated the idea of the $P$-violation in weak interactions. The
mirror matter sector (called also the ``hidden sector'') of field theory restores the
equivalence of the processes in the universe under the mirror reflection. Later the mirror
matter concept was developed for the $CP$-violation by Kobzarev, Okun and Pomeranchuk  \cite{KobOkuPom66} (see
\cite{Oku07} for the historical retrospective and complete bibliography).

Briefly, the concept of mirror matter is as follows. Space reflection operator $\vec r\to-\vec
r$ corresponds to the inversion operator $I_r$ in the quantum particle space. The commutativity
with a time-translation $[H, I_r]=0$ implies the conserved quantum number, corresponding to
$I_r$. The Landau form of the operator $I_r=CP$, where $C$ is the charge-conjugation, is not
satisfactory due to existence of the $CP$-violation. Alternative representation in the Lee-Yang
form $I_r=PR$ leads to the concept of mirror matter, where operator $R$ translates particles
into the mirror ones. In result, $I_r\Psi_L=\Psi'_R$ and $I_r\Psi_R=\Psi'_L$, where the prime
denotes the mirror particles, and $R$ and $L$ are, respectively, the right and left components
of the wave functions. These components are defined by the corresponding chirality operators
$P_L$ and $P_R$. For example, $P_L=(1-\gamma^5)/2$, where $\gamma^5\equiv
i\gamma^0\gamma^1\gamma^2\gamma^3$ and $\gamma^i$ are Dirac matrices. The Lagrangian of the
Standard Model contains only the left-handed components of spinors $\Psi_L=P_L\Psi$. The
chirality operator is different from the helicity operator, defined as
$\vec\sigma\cdot\hat{\vec p}/2 $, where $\hat{\vec p}=\vec p/p$, $\vec\sigma=
(\sigma_1,\sigma_2,\sigma_3)$, and $\sigma_i$ are the Pauli matrices. Both chirality and
helicity of particles are changed under the mirror reflection. The helicity operator tends to
the chirality operator only in the high-energy limit, $E\gg m$.

Nowadays, wormholes and mirror particles are the challenging problems in theoretical physics.
Different observational signatures of wormholes and mirror particles were proposed and
elaborated. Neutrino mixing and oscillations between the mirror and our world were considered
in \cite{AkhBerSen92,BerMoh95,BerVil00,BerNarVis03}. The reason for these oscillations is the
local gravitational interaction between our and mirror particles (the interaction terms in the
Lagrangian). This type of the gravitational interaction between the ordinary and mirror sectors
was proposed in \cite{KobOkuPom66}. See also \cite{BerCom01,Beretal05,BenBer01,Ber04} and
references therein for quantitative cosmological and astrophysical aspects of the mirror
matter. The passage of particles through wormholes is discussed in \cite{Gon97,Pop10}.
Properties of the non-orientable solutions with a topology of the Klein bottle were studied in
\cite{GonAlo11}.

In this article we argue that any particle is transformed into a corresponding mirror particle
and vice versa while moving through the non-orientable wormhole. This idea was first proposed
by Ya.~Ze'ldovich and I.~Novikov \cite{ZelNov67} for the universe with a general non-orientable
topology. Quite a similar idea was discussed later in
\cite{Sil01}. It was stated in \cite{HawIsr79} that  particle is transformed into its
antiparticle while moving through the non-orientable wormhole, but but process is difficult to
justify in view of the CP violation. Non-orientable wormholes are discussed in the book
\cite{Vis95} (p. 285), where it was pointed out that non-orientable wormholes are not
compatible with the Standard Model of elementary particles. Meainwhile, this problem is absent,
if particles, passing through the non-orientable wormhole, are transformed not into the
anti-particles of the Standard Model, but into the corresponding mirror particles. Note, that
transformation of the ordinary particle into the mirror one and vice versa is analogous to the
winding transformation around the ``Alice string'' \cite{Sch82,SchTyu82}.

In principle, the non-orientable wormholes may be the space-non-orientable, time-non-orientable
or even space-time-non-orientable \cite{Vis95}. We consider here only the space
non-orientability, which is related with the $P$ operation. The mirror matter sector, if it
exists, changes only the sense of some $CPT$ operations under the mirror reflection and does
not violate the $CPT$-theorem. Namely, in the mirror-matter models the operator $P$ is
determined up to the internal symmetry operator $R$, which transforms any particle from the
ordinary to mirror one, and vice versa. For this reason $PR$ is conserved. The $R$ operation
differs from the $C$ because the $CP$ is not an unbroken symmetry, and the nature is not
$CP$-invariant. By this reason, in the mirror matter models any particle, while travelling
around the non-orientable space, transforms not to its antiparticle but to the mirror particle.

The most direct astrophysical observational signature of the existence of non-orientable
wormholes, besides the gravitational effects, is a possible excess of matter in the vicinity of
the wormhole throats, ejected from the other side of wormholes with the mirror--ordinary
transformations. In particular, it is interesting to explore the idea of the mirror matter as a
candidate for the dark matter of the universe. In fact, the dark matter in the form of the
mirror matter particles was widely discussed and elaborated in details in many papers (see,
e.\,g., the discussion of different aspects of the mirror dark matter in the recent review
paper \cite{KusRos13}).

The mirror particles have their own mirror electric charge and photons with a corresponding
mirror electromagnetic interactions. Therefore, during traversing the non-orientable wormhole
throat, the ordinary charge disappear, but, instead of, the wormhole throat becomes charged.
The winding up of the electric field lines around the wormhole throats were considered by
J.~A.~Wheeler with relation to the ``charge without charge'' concept \cite{Whe55},
\cite{Whe62}. In the case of the non-orientable wormhole, the ordinary-to-mirror transformation
will also alternate the field lines from the ordinary to the mirror ones, as we discuss in this
paper.

The paper is organized as following. In the Section~\ref{norisec} we present the mathematical
description of the non-orientable wormholes. In the  auxiliary Section~\ref{hydsec} we
demonstrate the hydrodynamical traversability through some definite types of wormholes. This
traversability is needed for the justification of the baryons flows with mirror--ordinary
transformations, considered as a particular example in the Section~\ref{baryonsec}. When in the
Section~\ref{electrosec} we discuss non-trivial aspects of the non-orientable wormholes
electrodynamics in the presence of the mirror--ordinary charged particles transformations.
Finally, in the Section~\ref{conclsec} we write some conclusions.


\section{Non-orientable wormholes}
\label{norisec}

It was indicated in \cite{ZelNov67} that any particle, after traversing the universe with a
non-orientable topology on a certain spatially closed trajectory, can turn into the mirror one.
The very specific representation of a non-orientable space is the non-orientable wormhole
\cite{Vis95}, which can be constructed as a simple generalization of the ordinary wormhole
discussed in details by \cite{MorTho88-2}. Let us write the most general form of the
spherically symmetric static metric:
\begin{equation}
\label{metric} ds^2=f_1(x)dt^2 - f_2(x)^{-1}dr^2 - r^2(d\theta^2+\sin^2\theta d\phi^2),
\end{equation}
where $x=r/M$, $M$ is a mass parameter, and $G=c=1$ units are used. For example, in the case of
a charged Reissner-Nordstr\"om black hole
\begin{equation}
f_1(x)=f_2(x)=1-\frac{2}{x}+\frac{e^2}{x^2},
\end{equation}
where $e=Q/M$ is an electric charge of the black hole in the dimensionless units. The wormhole
metric is often written in the form \cite{MorTho88-2}:
\begin{equation}
f_1(x)=e^{2\Phi(x)},
\end{equation}
\begin{equation}
f_2(x)=1-\frac{K(r)}{r}=1-\frac{S(x)}{x},
\end{equation}
where the dimensionless function $S(x)$ is introduced. The $S(x)$ is related to the
(Misner-Sharp or Hawking) mass function. Some specific substance, violating the weak energy
conditions, is required for the construction of the wormhole throat to support a wormhole in
the steady state. For example, the phantom energy is used for this construction \cite{Sus05}.

\begin{figure}[t]
\begin{center}
\includegraphics[angle=0,width=0.49\textwidth]{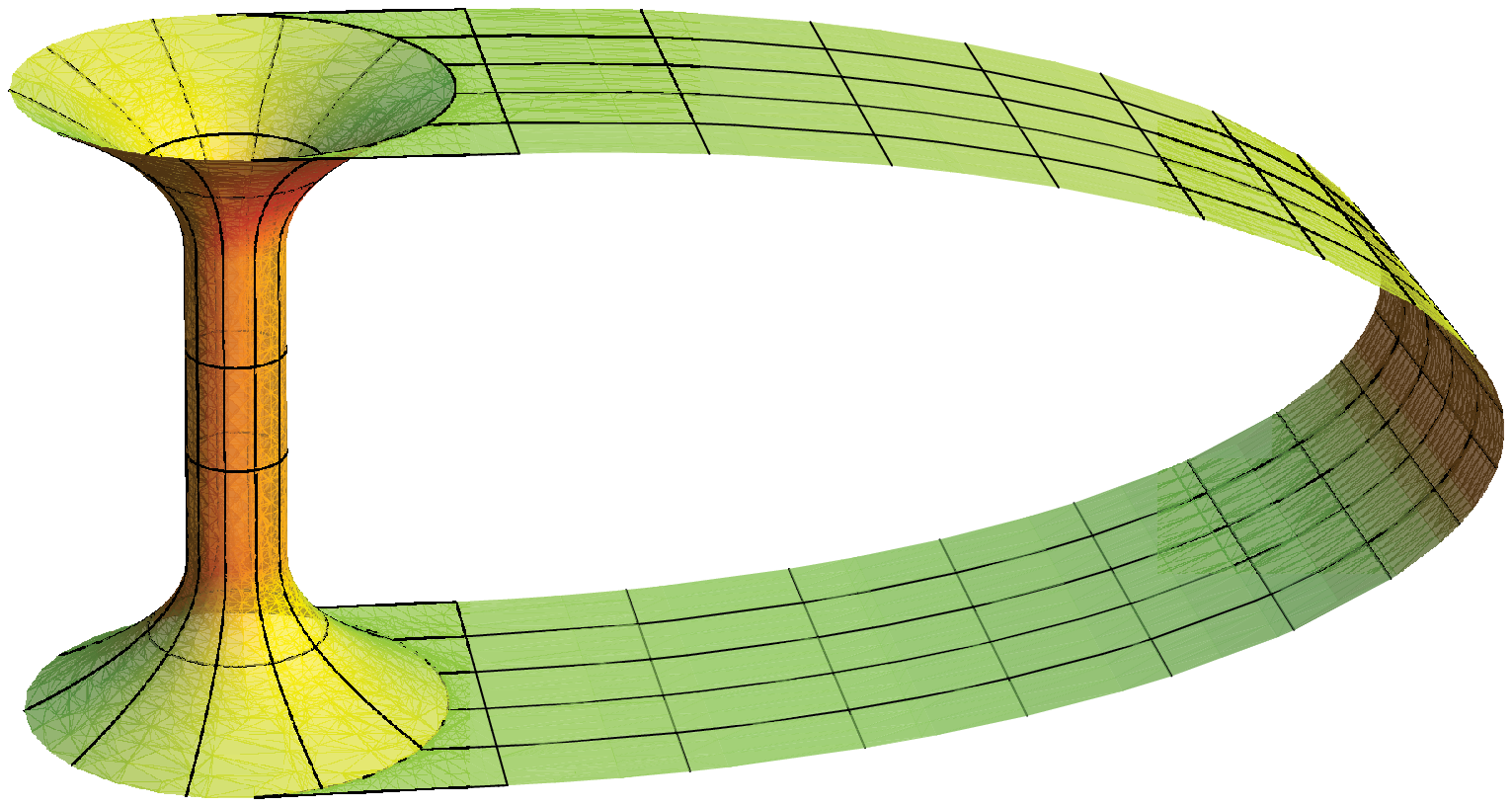}
\includegraphics[angle=0,width=0.49\textwidth]{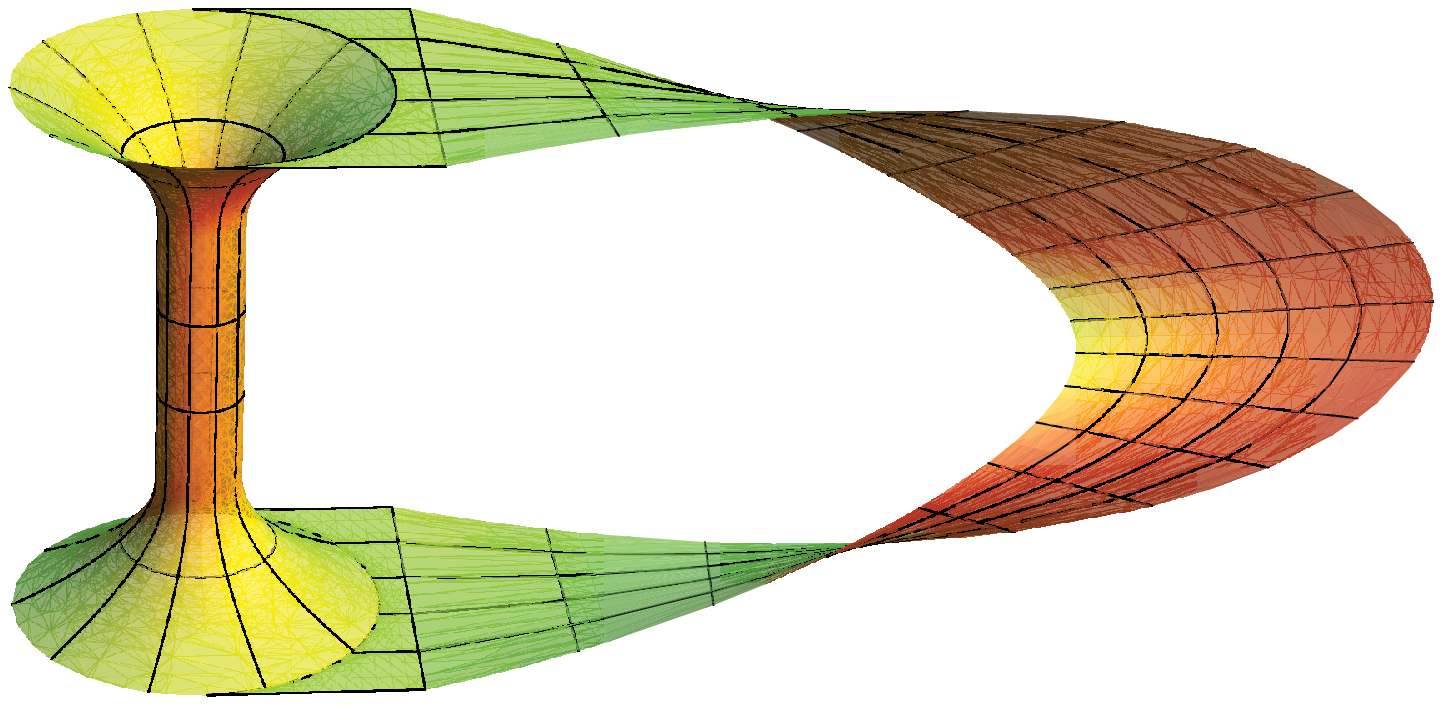}
\end{center}
\caption{Ordinary and non-orientable wormholes.} \label{gr2}
\end{figure}

In the oriented spaces, a smooth gluing of the angular coordinates at the edges of metric map
produces a total geometry with the globally fixed orientation. In the case of wormholes, the
gluing of separate maps produced at the throats $r=K(r)$:
\begin{equation}
\theta'=\theta, \qquad \phi'=\phi. \label{plus}
\end{equation}
This is illustrated in Fig.~\ref{gr2} (see the left panel), where only a spatial part of the
global wormhole geometry is qualitatively presented. However, in principle the inversion of
orientation is possible in the form
\begin{equation}
\theta'=-\theta, \qquad \phi'=\phi. \label{minus}
\end{equation}
This inverse gluing of the corresponding maps is similar to the construction of the M\"obius
strip. In the wormhole case, when there are two throats in the same universe, the inverse
gluing produces a non-orientable wormhole (see the right panel in Fig.~\ref{gr2}). Therefore,
any wormhole solution with the both throats in our universe can be transformed into the
non-orientable one by the inverse gluing of space coordinates at the throat. While considering
the ordinary and mirror particles in the same space-time, we may describe them for simplicity
and obviousness on the opposite sides of the sheets in Fig.~\ref{gr2}, which are, respectively,
the ordinary an mirror worlds. Note that a similar procedure of the inverse gluing produces the
Reissner-Nordstr\"om or Kerr black hole with an inverse orientation of the internal universes
inside the corresponding inner horizons.

The calculations in the non-orientable spaces require some caution. For example, in the curved
space-time case the $\gamma^5$ operator is changed as (see, e.\,g., \cite{Vis95},p. 286)
\begin{equation}
\gamma^5\equiv i\frac{\varepsilon_{[\mu\nu\lambda\rho]}}{4}
e^{\mu}_{\phantom{00}(a)}e^{\nu}_{\phantom{00}(b)}e^{\lambda}_{\phantom{00}(c)}e^{\rho}_{\phantom{00}(d)}
\gamma^{(a)}\gamma^{(b)}\gamma^{(c)}\gamma^{(d)},
\end{equation}
The chirality operator is changed in a similar way. In addition, the direct using of the Gauss
theorem reveals the non-trivial electrodynamic properties of non-orientable wormholes, which
will be discussed in the Section~\ref{electrosec}.


\section{Hydrodynamical traversability of wormholes}
\label{hydsec}

We describe here the hydrodynamical flows of matter between the mirror and our worlds through a
wormhole throat. We will use the approximation of the perfect fluid moving in the background
metric. The corresponding energy-momentum tensor of the perfect fluid
\begin{equation}
\label{emt} T_{\mu\nu}=(\rho+p)u_\mu u_\nu - pg_{\mu\nu},
\end{equation}
where $\rho$ and $p$ are, respectively, the energy density and pressure of the fluid in the
rest frame, $u^\mu=dx^\mu/ds$ is the 4-velocity. The adiabatic equation of state $p=p(\rho)$ is
assumed. For the wormhole metric we use an exact solution \cite{Sus05} with the phantom energy
supporting the wormhole throat. The phantom energy in this model is concentrated in the finite
region $r_0<r<r_1$ between the throat $r=r_0$ and some radius $r=r_1$. For $r>r_1$ the wormhole
metric coincides with the Schwarzschild black hole metric. The influence of the wormhole on the
matter at $r>r_1$ can be considered similar to the black hole case. We are interesting in the
possibility of the stationary accretion flow through the wormhole throat. A stationary
accretion is possible, if there is a critical sonic point in the flow. The accretion of perfect
fluid on the wormhole was discussed in \cite{Gon05} without the reference to the critical
point. Note, that a direction of the accretion flow (inflow or outflow) for individual wormhole
will depend on the initial conditions in the ambient plasma in the vicinity of the wormhole.

We consider the stationary accretion in the approximation of the test fluid, following the
general method of Michel \cite{Mic72}. We define an auxiliary function $dn/n=d\rho/(\rho+p)$,
which corresponds to the number density of particles in the case of the atomic gas. We solve
equations of motion $T^{\mu\nu}_{\; \; \;; \nu}=0$ in the given background metric. From a zero
$\mu=0$ component of these equations, and from the projection equation
$u_{\mu}T^{\mu\nu}_{\quad;\nu}=0$, it follows the two integrals of motion:
\begin{equation}
 \label{eq1}
  (\rho+p)\left(f_2+u^2\right)^{1/2}\gamma^2x^2 u =C_1,
\end{equation}
\begin{equation}
 \label{eq2}
nu\gamma x^2=-An_{\infty},
\end{equation}
where $u\equiv u^1$, $\gamma(x)\equiv(f_1/f_2)^{1/2}$, and $n_{\infty}$ is the value of $n$ at
$x\to\infty$. The rate of mass flow toward the coordinate center, $\dot M=-4\pi r^2T_0^{\;r}$,
is expressed in the form \cite{weprl}:
\begin{equation}
 \label{evol}
 \dot{M}=4\pi A M^2 [\rho_{\infty}+p(\rho_{\infty})].
\end{equation}
The numerical constant $A$ can be found by the fixing of the mass flow rate at the critical
sound point, and usually $A\sim{\cal O}(1)$. To calculate the fluid parameters at the critical
point, it is convenient to divide (\ref{eq1}) into (\ref{eq2}). This procedure gives:
\begin{equation}
 \label{eq3}
  (\rho+p)\left(f_2+u^2\right)^{1/2}\gamma n/n_{\infty} =C_2,
\end{equation}
where $C_2\equiv-C_1/A $. We denote by $V^2=dp/d\rho$ the square of the sound speed of the
accreted fluid. The critical point is calculated by the standard method from \cite{Mic72}. We
take the logarithmic differential of (\ref{eq2}) and (\ref{eq3}), exclude $dn/n$ and put the
multipliers of $dx/x$ and $du/u$ to zero. In this way, we find the system of equations for the
critical point:
\begin{equation}
V^2=\frac{u^2}{f_2+u^2}
 \label{sys1}
\end{equation}
\begin{equation}
2V^2\left(2+\frac{\gamma'x}{\gamma}\right)=2\frac{\gamma'x}{\gamma}+\frac{f_2'x}{f_2+u^2},
 \label{sys2}
\end{equation}
where `primes' denote the derivatives with respect to $x$. It is easy to check that in the
case, when $f_1$ and $f_2$ correspond to the Reissner-Nordstr\"om black hole metric, this
system of equations yields the known parameters of the critical point \cite{bde11}.

Let us now consider the critical point for the wormhole. Recall, that we consider the test
fluid without a back-reaction on the metric. Gravity of the accreted liquid would lead to some
shift of both the metric parameters and the critical point position. From the system of
equations (\ref{sys1}), (\ref{sys2}), we find the square of the fluid velocity at the critical
point:
\begin{equation}
u^2=\frac{1}{2}\Phi'(x-S).
 \label{v2crit}
\end{equation}
It is clear a stationary accretion takes place in the case of finite $\Phi'$ and $x>S$ (outside
the throat). However, the fluid velocity at $x=S$ should not be zero, while the fluid passes
through the throat. The stationary accretion condition is possible only if
$\Phi'\propto(x-S)^{-1}$, and $\Phi'$ should remain positive. This condition is satisfied for
the wormhole supported by the phantom energy \cite{Sus05}. Indeed, the corresponding Einstein
equations were written in \cite{Sus05} for the wormhole metric. In our notations, the
corresponding equations have the form:
\begin{equation}
S'=8\pi x^2\tilde\rho M^2,
 \label{ein1}
\end{equation}
\begin{equation}
\Phi'=\frac{8\pi x^3 \tilde p M^2+S}{2x(x-S)},
 \label{ein2}
\end{equation}
where $\tilde\rho$ and $\tilde p$ are, respectively, the energy density and pressure of the
supporting phantom energy. It is clear from (\ref{ein2}), that condition
$\Phi'\propto(x-S)^{-1}$ is valid for $x\to S$. Note, that the a similar traversable wormholes,
supported by the phantom energy, were considered also in \cite{GarLob07}.

A radial position $x$ of the critical point obeys the following equation
\begin{equation}
V^2(2+\Phi'x)=\Phi'x.
 \label{xcrit}
\end{equation}
By substituting $\Phi'$ from (\ref{ein2}), we obtain the following equation, which is non-linear in general,
\begin{equation}
V^2[4(x-S)+8\pi x^3 \tilde p M^2+S]=8\pi x^3 \tilde p M^2+S.
 \label{eqcp1}
\end{equation}
It's immediately seen that $x=S=x_0$ obeys this Equation, where $x_0=(-8\pi\tilde p_0 M^2)^{-1/2}$ is a throat radius, and $p_0$ is a pressure of the
phantom energy at the throat. Therefore, one of the hydrodynamical sonic point lies exactly on the throat of the wormhole.

If a near uniform distribution of the supporting phantom energy is realized in the vicinity of the
throat \cite{Sus05}, the solution of equation (\ref{ein1}) in the vicinity of the throat is
\begin{equation}
S=x_0+8\pi\tilde\rho M^2(x^3-x_0^3)/3.
\end{equation}
Respectively, equation (\ref{eqcp1}) in this case takes the form
\begin{equation}
y^3\left(\frac{1-V^2}{1+3V^2}+\frac{1}{3w}\right)+\frac{4V^2}{1+3V^2}y-\frac{1+3w}{3w}=0
\label{eqcube}
\end{equation}
where $y\equiv x/x_0$, $w\equiv\tilde p/\tilde\rho<-1$. It is easy to see that $y_1=1$ is the
root of equation (\ref{eqcube}) for any $V$ and $w$. The other two roots can be found from the
Vieta relations:
\begin{equation}
1+y_2+y_3=0, \qquad y_2y_3=-\frac{1+3w}{3w} \label{viet}
\end{equation}
in the form
\begin{equation}
y_2=-\frac{1}{2}\pm\frac{1}{2}\sqrt{\frac{4+15w}{3w}}, \qquad
y_3=-\frac{1}{2}\mp\frac{1}{2}\sqrt{\frac{4+15w}{3w}}. \label{y12}
\end{equation}
One of the roots, $y_{2,3}$, could be $\geq1$ only for $0<w\leq1/3$. However, in the considered
phantom case with $w<-1$, the roots are $<1$ or even complex. This means that solution with
$y<1$ does not correspond to any critical (sonic) point and, therefore, is non-physical. In
result, we demonstrate that there is a critical point in the fluid flow, which is located exactly at the throat of the wormhole. This means that the described wormhole is traversable for the perfect fluid. There are may be another sonic points outside the 
phantom energy region, where the Schwarzschild solution is valid, see \cite{weprl}. In this case the fluid must flow subsequently through all the 
sonic points. This is the condition of the stationary self-consistent solution.

At the second throat of the wormhole, a fluid flow would have the same stationary density
distribution as in the accretion flow toward the first throat, but with the reversed direction
of flow, i.\,e., with the outward directed fluid velocity $u>0$.

\section{Accretion of baryons by the non-orientable wormhole}
\label{baryonsec}

Astrophysical signatures of the mirror matter in the Galaxy were considered in \cite{BliKhl83},
where the gravitational influence of the mirror matter on the clustering of ordinary baryons
were investigated. We consider now the physical effect, which may be the observational
signature of the existence of non-orientable wormhole. The difference between the ordinary
matter and the mirror one is related with the asymmetry of the inflation potential, see,
e.\,g., \cite{BerDolMoh96}, \cite{BerVil00}. As an example, we will assume that mirror matter
is the dark matter in the our universe \cite{KusRos13}. The dark matter density in the Galactic
halo is a rather well described by the Navarro-Frenk-White profile \cite{NavFreWhi96}
\begin{equation}
 \rho_{\rm H}(r)=
 \frac{\rho_{0}}{\left(r/R_s\right)\left(1+r/R_s\right)^2},
 \label{halonfw}
\end{equation}
where $R_s=20$~kpc is a scale radius, $R_h=$~200~kpc is a virial radius of the halo, $\rho_{\rm
h}(r_\odot)=0.3$~GeV~cm $^{-3}$ is an average density at the distance of the Sun,
$R=r_{\odot}=8.5$~kpc, from the Galactic center. The baryonic matter dominates at the distances
$r\le r_\odot$. In this model the Navarro-Frenk-White profile gives the density of mirror
matter.

If the density and pressure of matter in the one chosen world, our or mirror one, is much
greater than the corresponding values in the other world, the collision of the
counter-propagating streams in the wormhole throats can be ignored. Under this condition, there
will be a net flow from the world with a greater density into the world with a lower density
through the wormhole throats. For example, let us consider the case, when the first throat is
located in the halo of our Galaxy far beyond the stellar disk at some distance $r=r_w$ from the
Galactic center, and the second throat of the same whormhole is located at the place, where the
baryon density is $\ll\rho_{\rm H}(r_w)$. In this case, the mirror matter will flow through the
wormhole and will turn into the ordinary one at the second throat. In result, the region of the
moving matter with an unusual chemical composition will be observed near the second throat.

We suppose that mirror dark matter forms the major part of the extended Galactic halo in the
our world. For simplicity, we suppose also that a mean density of the mirror gas in the
Galactic halo is of the same order as a mean density of the mirror stars. The heavy elements
define the size of the forming baryonic halo of Galaxy, because the epoch of the halo
stabilization depends on the star formation history, which, in turn, strongly depends on
ambient chemical composition. This means that the mirror matter is stabilized at the Galactic
halo outskirts, avoiding contraction into the disk. As a consequence, the heavy elements would
be more abundant in the mirror universe, because the mirror stars are formed more efficiently
in the halo. A similar idea, underlying the specific properties of the mirror world,  is
elaborated  in \cite{zurabas}.

As an illustrative example, we consider the Bondi-Hoyle accretion onto the wormhole with a mass
$M$ located at the distance $l=30$~kpc from the Galactic center, where the local density
(\ref{halonfw}) is $\rho_{\rm H}\simeq5.7\times10^{-26}$~g~cm$^{-3}$. The mass rate of the
Bondi-Hoyle accretion, i.\,e., the rate of transformation of the mirror matter into the our
baryonic matter is
\begin{eqnarray}
\dot M&=&\frac{4\pi G^2M^2\rho}{v^3} \label{bondi}
\\
&\simeq&1.6\times10^{10}\left(\frac{M}{10^2M_\odot}\right)^2
\left(\frac{v}{200\mbox{~km~s$^{-1}$}}\right)^{-3}\mbox{~g~s$^{-1}$},\nonumber
\end{eqnarray}
where $v$ is a sound speed or the wormhole throat  velocity through the ambient gas. We choose
the virial velocity as a distinct and representative value. The accretion energy is radiated
outward if the throat radius is smaller than the accretion disk  radius. The corresponding
luminosity of this object is $L_X=\eta\dot Mc^2$, where $\eta\simeq0.1$ is a typical efficiency
of the accretion mass-energy conversion. For the normalization parameters used in
(\ref{bondi}), the corresponding luminosity is $L_X\sim1.4\times10^{30}$~erg~$^{-1}$, i.\,e.,
it is at the level of faint stars luminosity. This radiation is emitted in the mirror world,
passed partially through the wormhole and is emitted finally in the our world. The passage of
light through a wormhole was examined in \cite{Doretal09,Sha09}. Therefore, some non-identified
X-ray objects in the Galaxy can be related, in principle, with the mirror matter outflows from
the non-orientable wormholes.

\section{Electrodynamics of the non-orientable wormholes}
\label{electrosec}

The mirror-to-ordinary transformations of the charged particles and fields would modify also
the ``charge without charge'' concept, proposed by J.~A.~Wheeler \cite{Whe55,Whe62}. Let us
consider the ``charge without charge'' construction, where electric field lines are winded up
through two wormhole throats. The strength of the resulting global electric field could be very
strong near the wormhole throat without existence of any real charge, mimicking the violation
of the Gauss theorem. In fact, the standard formulation of the Gauss theorem is applicable only
to the closed or topologically simply connected space-time backgrounds. Meantime, the winding
problem for the electric (or magnetic) field  becomes nontrivial in the case of the
non-orientable wormholes and with the presence of mirror matter. In the ordinary case, i.\,e.,
without the mirror sector, any field line, that goes out of the throat, corresponds to the
ordinary electric field line. Respectively, the net ``charge without charge'' of the wormhole
throat is proportional to the number of winding loops of these lines. 

Quite the contrary, this
correspondence changes radically, if one takes into account the mirror-to-ordinary
transformation in the presence of the mirror sector with mirror electric charges and mirror
photons and if the wormhole is non-orientable. To elucidate this problem one needs to consider
the ``charge without charge'' concept with the electric charge $q$ moving repeatedly along the
winding up trajectory. Let us suppose that we see the ordinary electric charge, which several times enters the left throat and exits from the right throat. After the first
passage through the non-orientable wormhole, the charge becomes the mirror one. 
After the second passage the charge becomes the ordinary, and so on. The effective electric charge (``charge without charge'') of the throat can
be expressed through the electric flow as usual
\begin{equation}
Q=\frac{1}{4\pi}\oint {\bf E}d{\bf S}.
 \label{flow}
\end{equation}
As a result, in the case of the
non-orientable wormhole, we will see only half of the ``charge without charge'' in comparison
with the net charge of the corresponding orientable wormhole. More exactly, after the $n$ passages the left and 
the right throats will have the charges
\begin{equation}
Q_l=q\left[\frac{n+1}{2}\right],\qquad Q_r=-q\left[\frac{n}{2}\right],
 \label{qlqr}
\end{equation}
where $[...]$ is the integer part of a number. This is the specific feature of the non-orientable wormhole electrodynamics. Note that similar conclusion is true for the lepton, baryon and other conserved quantum charges.

Now we consider the following thought experiment. Let there are two close Coulomb interacting charges $q_1$ and $q_2$. Lets the $q_2$ went 
through the non-orientable wormhole and approached the $q_1$ again. The $q_2$ is now the mirror charge, so the direct interaction through the short line is impossible. But there is still the long way through the non-orientable wormhole. The charges can exchange by virtual photons (they transform into the mirror photons on the way) through the non-orientable wormhole and can experience some interaction through the long way. By other words the wormhole's throat becomes electrically charged $q=q_2$, and the $q_1$ interacts with the charged wormhole's throat. If the charges are not far from the wormhole's throat, the throat may not be the point charge, so the charge $q_1$ can feel the small-scale structure of the field, while it ``sees'' the charge $q_2$ through the wormhole's throat. In another thought experiment both the charges $q_1$ and $q_2$ fly together through the non-orientable wormhole. The fundamental constants values are the same in our and mirror worlds, therefore the charges will continue to interact one with another with the permanent strength all the time. But before the fly by the interaction is mediated by the ordinary electric field, and after the fly by the interaction is mediated by the mirror field. Note also that there is no any ``switching'' point along the way, because the ordinary or mirror nature of the mediated forces is the only question of the side of view.

\section {Conclusion}
\label{conclsec}

In this paper we consider the relations between the non-orientable wormholes and mirror
particles. The main idea of our paper is in the fusion of the non-orientable wormholes and the
mirror matter concept. In particular, it is stated that the mirror particles are transformed
into the ordinary ones and vice versa while traversing through the non-orientable wormhole.
This statement was previously applied only to the universe with a non-orientable topology
\cite{ZelNov67}. Here this concept is applied to a particular solution of the Einstein
equations in the form of the non-orientable wormhole. In the framework of this hypothetical
solution we expect the existence of matter flows through the wormhole throats, accompanied by
the transformations between ordinary and mirror particles. This nonlocal or topological
transformation is additional to the local gravitational interaction between the ordinary and
mirror matter.

We described the hydrodynamical traversability of the wormholes by calculating the stationary
accretion flow of perfect fluid through the wormhole throat. The wormhole, supported by the
phantom energy, appears to be the traversable for this hydrodynamics flows. As the possible
astrophysical signatures of the non-orientable wormhole hypothesis we suggest the observation
of mass fluxes and the non-identified X-ray sources, assuming that the dark matter in the
universe is composed of the mirror matter.

As an alternative to the non-orientable wormhole, one can consider the Reissner-Nordstr\"om or
Kerr black hole global geometry with a modified connection between the external and internal
universes similar to the M\"obius strip (\ref{minus}). The resulting non-orientable black hole
would connect our universe with the internal mirror universe. Matter inflowing through this
black hole would be converted into the mirror matter traversing to the internal universe.
However, this is only the one way flow without the possibility to return back. In principle, we
may see these outflows of mirrir matter, if there are white holes in the our universe.

\section*{Acknowledgments}
Authors acknowledge Z. Berezhiani and A. Gazizov for the enlightening discussions of the mirror
matter problems. This research was supported in part by the grants NSh-3110.2014.2 and
RFBR~13-02-00257-a.

\end{document}